\documentclass[aps,twocolumn,groupedaddress,showpacs]{revtex4}
\usepackage{graphicx}

\begin{document}

\title[]
{Phase diagram of the 3D Axial-Next-Nearest-Neighbor Ising model}

\author{A.~Gendiar$^{1,2}$ and T.~Nishino$^2$}
\affiliation{$^1$Institute of Electrical Engineering, Slovak Academy of Sciences,
D\'ubravsk\'a cesta 9, SK-842 39 Bratislava, Slovakia\\
$^2$Department of Physics, Faculty of Science, Kobe University,
Kobe 657-8501, Japan}

\date{\today}

\begin{abstract}
The three-dimensional axial-next-nearest-neighbor Ising (ANNNI) model
is studied by a modified tensor product variational approach (TPVA).
A global phase diagram is constructed with numerous commensurate and
incommensurate magnetic structures. The devil's stairs behavior for
the model is confirmed. The wavelength of the spin modulated phases
increases to infinity at the boundary with the ferromagnetic phase.
Widths of the commensurate phases are considerably narrower
than those calculated by mean-field approximations.
\end{abstract}


\maketitle

\section{Introduction}

Periodically modulated magnetic structures have attracted scientific interest
for several decades both experimentally and theoretically. A non-trivial phase
diagram obtained by experimental measurements in cerium antimonide (CeSb)
shows a variety of different commensurately ordered magnetic structures with
the underlying lattice~\cite{CeSb1,CeSb2}. The three-dimensional (3D) $S=\frac{5}{2}$
axial next-nearest-neighbor Ising (ANNNI) model has been considered as a theoretical
candidate for CeSb since it exhibits a rich structure when it is treated by
mean-field approximation~\cite{Boehm}. The 3D $S=\frac{1}{2}$ ANNNI model is
another example that shows a non-trivial spin modulated phase --- the so-called
devil's stairs. This model has been analyzed theoretically by various approaches,
including high-temperature series expansions~\cite{Redner,Otmaa}, low-temperature
series expansions~\cite{FischerSelke}, mean-field approximations~\cite{MFA},
Monte-Carlo simulations~\cite{MCS}, an effective-field approximation~\cite{Surda}, 
free-fermion methods, a phenomenological renormalization, and
other methods reviewed in Ref.~\cite{RepBak,SelkeRev}.
The Monte Carlo simulations have also been applied to the $S=\frac{1}{2}$ ANNNI
model with a finite number of spin layers~\cite{Selke}. Recently, Henkel and Pleimling
considered an anisotropic scaling at the Lifshitz point using the Wolff cluster
algorithm and critical exponents have been calculated~\cite{Henkel}.

The purpose of this paper is to clarify the phase structure of the 3D
$S=\frac{1}{2}$ ANNNI model. Our interest is to study the spin modulated
phases at intermediate temperatures, particularly, the stability of
commensurate phases. For this purpose we apply a numerical variational
method, the tensor product variational method (TPVA), to the model.
In Section~II we introduce the 3D ANNNI model and briefly discuss the variational
background of the TPVA applied to the system. We present the numerical
results in Section~III where we construct the global phase diagram
of the model and analyze the spin modulated phases. We summarize the obtained
results in Section~IV. In Appendix, a numerical self-consistent optimizing
process is reviewed and efficiency of the modified TPVA is discussed.


\section{Model and Non-uniform Product Variational State}

We study the $S=\frac{1}{2}$ ANNNI model on a simple cubic lattice with
the size $L\times\infty\times\infty$ along the $x$, $y$, and $z$ directions,
respectively. The model is described by the lattice Hamiltonian
\begin{eqnarray}
\nonumber
\label{Hamil}
{\cal H}&=&-J_1\sum\limits_{i,j,k}\sigma_{i,j,k}\left(
\sigma_{i+1,j,k}\,+\,\sigma_{i,j+1,k}\,+\,\sigma_{i,j,k+1}\right)\\
&&+J_2\sum\limits_{i,j,k}\sigma_{i,j,k}\ \sigma_{i+2,j,k}\ ,
\end{eqnarray}
where the subscripts $i$, $j$, and $k$ of the Ising spin $\sigma=\pm1$ refer to
the $x$, $y$, and $z$ coordinates, respectively. The ferromagnetic interaction
$J_1>0$ acts between the nearest-neighbors and $J_2>0$ is the competing
antiferromagnetic interaction between the next-nearest-neighbors imposed
only in the $x$ direction.

\begin{figure}[tb]
\includegraphics[width=85mm,clip]{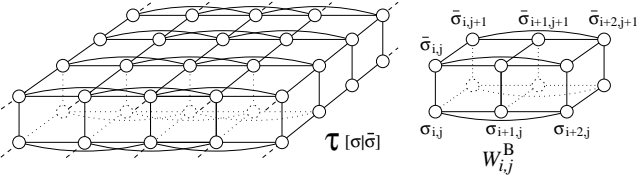}
\caption{The layer-to-layer transfer matrix ${\cal T}[\sigma|\bar\sigma]$
(left) illustrated in the case for $L=5$ and the local Boltzmann weight
$W_{i,j}^{\rm B}\{\sigma|\bar\sigma\}$ (right).}
\label{bwtm}
\end{figure}

Figure~\ref{bwtm} shows the layer-to-layer transfer matrix ${\cal T}$ to
the $z$ direction which connects two adjacent spin layers $[\sigma]$ and
$[\bar\sigma]$ (each of the size $L\times\infty$ in the $x$ and $y$
directions). The transfer matrix can be exactly expressed as the product
of partially overlapped local Boltzmann weights (cf.~Fig.~\ref{bwtm})
\begin{equation}
{\cal T}[\sigma|\bar\sigma]=\prod\limits_{i=1}^{L-2}\ \ \prod\limits_{j=-\infty}^{+\infty}
W^{\rm B}_{i,j}\{\sigma|\bar\sigma\}\ .
\end{equation}
We simplify the notations using a group of 6 spins
\begin{equation}
\label{spin6}
\{\sigma\}\equiv(\sigma_{i,j}\ \sigma_{i^\prime,j}\ \sigma_{i^\prime{}^\prime,j}
\ \sigma_{i,j^\prime}\ \sigma_{i^\prime,j^\prime}\ \sigma_{i^\prime{}^\prime,
j^\prime})\ ,
\end{equation}
with the index rule $i^\prime=i+1$, $i^\prime{}^\prime=i+2$, and $j^\prime=j+1$.
The local Boltzmann weight $W^{\rm B}_{i,j}$ of the Hamiltonian in Eq.~(\ref{Hamil})
has the following form
\begin{eqnarray}
\label{bw}
W^{\rm B}_{i,j}\{\sigma|\bar\sigma\}=\exp\left\{\frac{1}{k_{\rm B}T}\left[\frac{J_1}{6}
\left(\sigma_{i,j}\bar\sigma_{i,j}+\sigma_{i^\prime,j}\bar\sigma_{i^\prime,j}\right.
\right.\right.\\
\nonumber
\left.
+\sigma_{i^\prime{}^\prime,j}\bar\sigma_{i^\prime{}^\prime,j}+\sigma_{i,j^\prime}
\bar\sigma_{i,j^\prime}+\sigma_{i^\prime,j^\prime}\bar\sigma_{i^\prime,j^\prime}
+\sigma_{i^\prime{}^\prime,j^\prime}\bar\sigma_{i^\prime{}^\prime,j^\prime}
\right)\\
\nonumber
+\frac{J_1}{8}\left(
\sigma_{i,j}\sigma_{i^\prime,j}+\sigma_{i^\prime,j}\sigma_{i^\prime{}^\prime,j}+
\sigma_{i,j^\prime}\sigma_{i^\prime,j^\prime}+\sigma_{i^\prime,j^\prime}
\sigma_{i^\prime{}^\prime,j^\prime}
\right.\\
\nonumber
\left.
+\bar\sigma_{i,j}\bar\sigma_{i^\prime,j}+\bar\sigma_{i^\prime,j}
\bar\sigma_{i^\prime{}^\prime,j}+\bar\sigma_{i,j^\prime}\bar\sigma_{i^\prime,j^\prime}
+\bar\sigma_{i^\prime,j^\prime}\bar\sigma_{i^\prime{}^\prime,j^\prime}
\right)\\
\nonumber
\left.\left.-\frac{J_2}{4}\left(
\sigma_{i,j}\sigma_{i^\prime{}^\prime,j}+\sigma_{i,j^\prime}\sigma_{i^\prime{}^\prime,j}
+\sigma_{i,j}\sigma_{i^\prime{}^\prime,j}+\sigma_{i,j^\prime}\sigma_{i^\prime{}^\prime,j}
\right)\right]\right\}
\end{eqnarray}
with $k_{\rm B}$ being the Boltzmann constant and the temperature $T$.
For reasons of simplicity and brevity, we consider $J_1$=1 and $k_{\rm B}$=1
throughout all calculations~\cite{rem1}.

\begin{figure}[tb]
\includegraphics[width=85mm,clip]{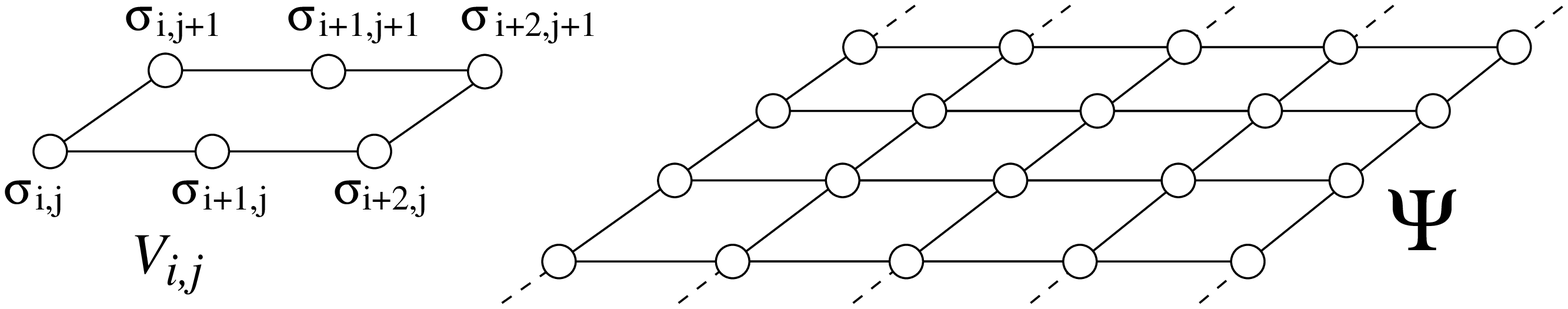}
\caption{Graphical representation of the local variational weight
$V_{i,j}\{\sigma\}$ (left) used to construct the trial function $\Psi$
(right) in the particular case for $L=5$.}
\label{lvwts}
\end{figure}

We consider a variational problem for the transfer matrix ${\cal T}[\sigma|\bar\sigma]$.
For a given trial state $|\Psi\rangle$, the variational partition function
per layer is given by
\begin{equation}
\lambda_{\rm var}(\Psi)=\frac{\langle\Psi|{\cal T}|\Psi\rangle}
{\langle\Psi|\Psi\rangle}=\frac{\sum\limits_{[\sigma],[\bar\sigma]}\Psi[\sigma]
{\cal T}[\sigma|\bar\sigma]\Psi[\bar\sigma]}{\sum\limits_{[\sigma]}(\Psi[\sigma])^2}.
\label{vpf}
\end{equation}
The TPVA is a numerical variational method that assumes a trial function
written by the product of local weights $V$. For the ANNNI model, $\Psi$ is
written in the product form of mutually overlapped local weights
(cf.~Fig.~\ref{lvwts})
\begin{equation}
\Psi[\sigma]
=\prod\limits_{i=1}^{L-2}\ \ \prod\limits_{j=-\infty}^{+\infty}V_{i,j}\{\sigma\},
\label{vs}
\end{equation}
where we have used the simplified notation in Eq.~(\ref{spin6}).

In order to study non-uniform spin modulated phases, the local variational
weights $V_{i,j}\{\sigma\}$ must be position dependent along the $x$ direction.
Each $V$ thus contains $2^6=64$ adjustable parameters. Since we have written
the trial function $\Psi$ as well as the transfer matrix ${\cal T}$ in
the product forms, both the numerator of Eq.~(\ref{vpf})
\begin{equation}
\langle\Psi|{\cal T}|\Psi\rangle=
\sum\limits_{[\sigma],[\sigma^\prime]}\ \ \prod\limits_{i=1}^{L-2}
\ \ \prod\limits_{j=-\infty}^{+\infty}V_{i,j}\{\sigma\}
W^{\rm B}_{i,j}\{\sigma|\bar\sigma\}V_{i,j}\{\bar\sigma\}
\label{eq1}
\end{equation}
and its denominator
\begin{equation}
\langle\Psi|\Psi\rangle=
{\sum\limits_{[\sigma]}\ \ \prod\limits_{i=1}^{L-2}
\ \ \prod\limits_{j=-\infty}^{+\infty}(V_{i,j}\{\sigma\})^2}
\label{eq2}
\end{equation}
have also the product forms. Thus these quantities can be accurately calculated
by means of renormalization techniques, particularly, we used the density matrix
renormalization group (DMRG)~\cite{White,Nishino}.


\section{Results}

\begin{figure}[tb]
\includegraphics[width=85mm,clip]{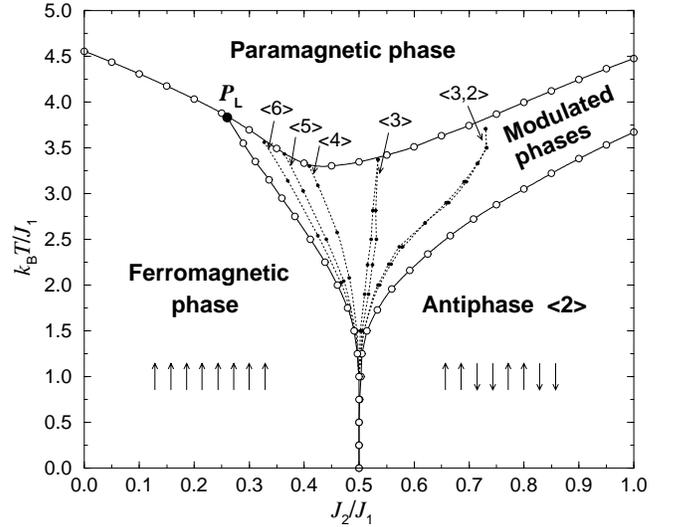}
\caption{The global phase diagram of the 3D ANNNI model obtained by the TPVA.
The Lifshitz point $P_{\rm L}$ is denoted by the black circle. The dotted
lines enclose extremely narrow commensurate phases.}
\label{ANNNIpd}
\end{figure}

Figure~\ref{ANNNIpd} shows the global phase diagram of the ANNNI model
obtained by the TPVA. It consists of
\begin{itemize}
\item[(i)] a paramagnetic (disordered) phase,
\item[(ii)] a uniformly ordered ferromagnetic phase,
\item[(iii)] an antiphase with the periodic spin alignment
($\cdots\uparrow\uparrow\downarrow\downarrow\cdots$)
for which we use the notation $\langle2\rangle$ in the following, and
\item[(iv)] a rich area of spin modulated phases.
\end{itemize}
Region of the spin modulated phases separates the antiphase from the
paramagnetic phase. The paramagnetic, ferromagnetic, and the modulated
phases meet at the Lifshitz point $P_{\rm L}$. In our calculations, it
is located at $J_2^{\rm L}/J_1$=0.26 and $k_{\rm B}T_{\rm L}/ J_1$=3.83
and is in agreement with the latest Monte Carlo calculations carried out
by Pleimling and Henkel $J_2^{\rm L}/J_1$=0.270(4) and
$k_{\rm B}T_{\rm L}/J_1$=3.7475(50) (from Ref.~\cite{Henkel}).

The resulting phase diagram does not contradict to previous knowledge of
the model. The phase boundary lines separating the ferromagnetic phase,
the paramagnetic phase, the antiphase, and the spin modulated phases
coincide with those obtained using the Monte Carlo calculations~\cite{MCS}.
We found new features of the model in the region of the modulated
phases, where the Monte Carlo simulations have not yielded a satisfactory
answer. Our results are thought of as a supplement to the achievements
computed by the mean-field approximations~\cite{MFA,Surda} at higher
temperatures and by the low-temperature series expansions valid at low
temperatures $k_{\rm B}T/J_1\ll4$~\cite{FischerSelke}.

\begin{figure}[tb]
\includegraphics[width=85mm,clip]{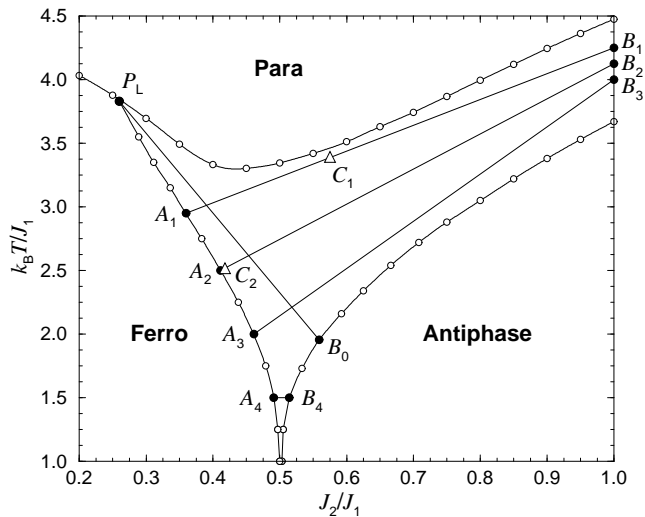}
\caption{The five selected lines $A_1B_1$, $A_2B_2$, $A_3B_3$, $P_{\rm L}B_0$,
and $A_4B_4$ in the region of modulated phases. The points $C_1$ and $C_2$ are
marked by the white triangles}
\label{mpl}
\end{figure}

In the rest of this section, we focus on the region of the modulated phases
that contains a multitude of various commensurate and incommensurate phases.
For example, in Fig.~\ref{ANNNIpd} we plotted a few narrow areas of typical
commensurate phases such as $\langle3,2\rangle=
(\cdots\uparrow\uparrow\uparrow\downarrow\downarrow\cdots)$, $\langle3\rangle=
(\cdots\uparrow\uparrow\uparrow\downarrow\downarrow\downarrow\cdots)$,
$\langle4\rangle=(\cdots\uparrow\uparrow\uparrow\uparrow\downarrow\downarrow
\downarrow\downarrow\cdots)$, etc. all enclosed by the dotted lines. Note
that the widths of these phases are substantially narrower compared to the
mean-field approximation~\cite{MFA} and the effective-field
approximation~\cite{Surda}.

\subsection{Wavelength analysis}
We first explain relation between the conventional notation and
and modulation wave length $\lambda$. The antiphase
$\langle2\rangle=\ \uparrow\uparrow\downarrow\downarrow
\uparrow\uparrow\downarrow\downarrow$ has periodicity of 4 lattice sites,
thus $\lambda=4$. Another examples is the high-order commensurate
phase $\langle3,(3,2)^2\rangle$ which represents the periodic spin sequence
$(\uparrow\uparrow\uparrow\downarrow\downarrow\downarrow\uparrow\uparrow
\downarrow\downarrow\downarrow\uparrow\uparrow)$ and yields $\lambda=26/5$.

\begin{table}[b]
\caption {The positions of the points depicted in Fig.~\ref{mpl}.}
\label{tb2}
\begin{ruledtabular}
\begin{tabular*}{\hsize}{
c@{\extracolsep{0ptplus1fil}}l@{\extracolsep{0ptplus1fil}}
l|@{\extracolsep{0ptplus1fil}}c@{\extracolsep{0ptplus1fil}}
l@{\extracolsep{0ptplus1fil}}l@{\extracolsep{0ptplus1fil}}}
{\ Point} & $J_2/J_1$ & $k_{\rm B}T/J_1$ \ \ & {\ Point} & $J_2/J_1$ & $k_{\rm B}T/J_1$  \\
\colrule
$P_{\rm L}$ & 0.2600 & 3.83   & $B_0$ & 0.555 & 1.9812 \\
$A_1$       & 0.3560 & 2.95   & $B_1$ & 1.000 & 4.2500 \\
$A_2$       & 0.4113 & 2.50   & $B_2$ & 1.000 & 4.1250 \\
$A_3$       & 0.4605 & 2.00   & $B_3$ & 1.000 & 4.0000 \\
$A_4$       & 0.4908 & 1.50   & $B_4$ & 0.514 & 1.5000 \\
$C_1$       & 0.5750 & 3.3921 & $C_2$ & 0.418 & 2.5185 \\
\end{tabular*}
\end{ruledtabular}
\end{table}

\begin{figure}[tb]
\includegraphics[width=85mm,clip]{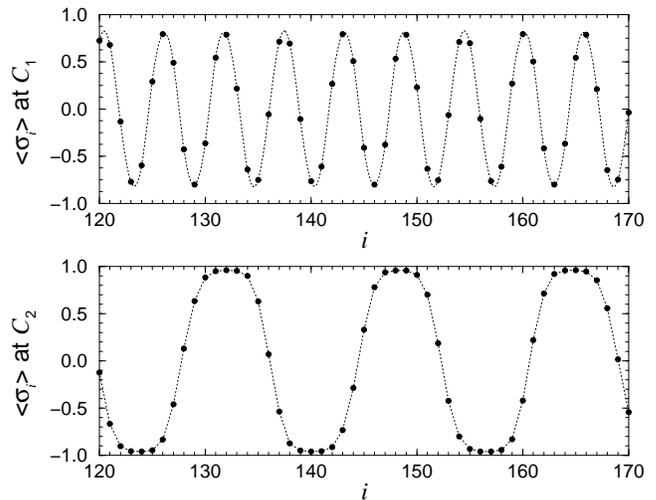}
\caption{The spontaneous magnetization $\langle\sigma_i\rangle$ versus
lattice size in the $x$ direction calculated at $C_1$ (the upper graph)
and at $C_2$ (the lower one) with the lattice size $L=401$. In order to
show the data in detail, we plot $i=120,\dots,170$.}
\label{c1}
\end{figure}

\begin{figure}[tb]
\includegraphics[width=85mm,clip]{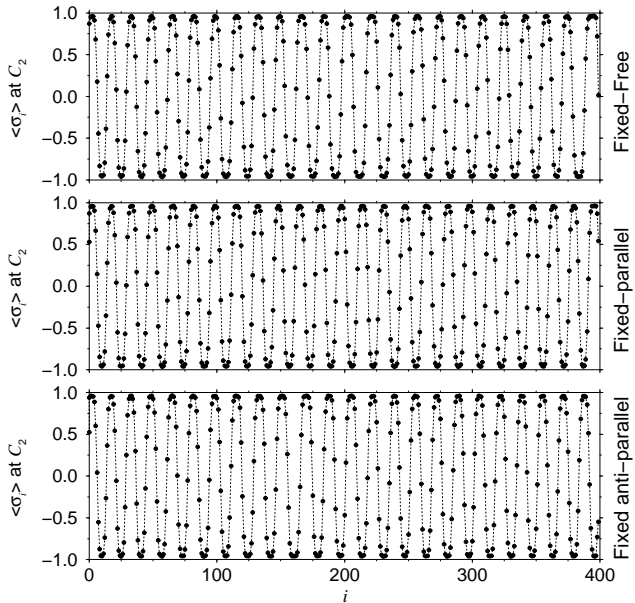}
\caption{The spontaneous magnetization obtained at $C_2$ for the lattice
size $401\times\infty\times\infty$. The upper, middle, and lower graphs
display $\langle\sigma_i\rangle$ for the three different boundary conditions.}
\label{c2}
\end{figure}

We calculate the spin modulations along five representative lines as
depicted in Fig.~\ref{mpl} with their ending points listed in Table~I.
When we obtain the spontaneous magnetization, we compute the corresponding
wavelengths by means of the Fourier transform.

Figure~\ref{c1} shows the spin polarizations $\langle\sigma_i\rangle$
at the two parameter points: $C_1$ on the line $A_1B_1$ and $C_2$ on $A_2B_2$.
These two points are chosen near the phase boundaries.
The spin polarization at $C_1$ exhibits the commensurate phase
$\langle3^5,2\rangle\equiv (\cdots\downarrow\downarrow\downarrow\uparrow
\uparrow\uparrow\downarrow\downarrow\downarrow\uparrow\uparrow\uparrow
\downarrow\downarrow\downarrow\uparrow\uparrow\cdots) $ with the wavelength
$\lambda=17/3$ (the upper graph). The lower graph shows the commensurate phase at
$C_2$ with $\lambda\approx16.7$ on the same region along the $x$ direction.

Now, we give a brief discussion on influence of boundary
conditions imposed to the system on the resulting spin polarization.
In Fig.~\ref{c2} we plot $\langle\sigma_i \rangle$ for three different
types of the boundary conditions. We consider a lattice with the size
$401\times\infty\times\infty$ and analyze the data at $C_2$. On the upper
graph, the spin polarization is calculated for the fixed boundaries on
the left end (the spins are aligned to the 'up' direction) and the free
boundaries on the right end. The Fourier transform applied to the whole
region $i=0,1,\dots,400$ yields $\lambda=16.7\pm0.53$. On the 
intermediate graph, the parallel fixed boundary conditions on both
sides are imposed (the spins are aligned 'up' at the ends). It gives
$\lambda=16.7\pm0.41$. Finally, the lower graph shows the anti-parallel
fixed boundaries ('up' on the left end and 'down' on the right end) with
$\lambda=16.7\pm0.41$. The choice of the boundary conditions does not
affect the numerical results significantly. The larger lattice size
is considered, the less influence of the boundaries is obtained, especially,
off of the phase boundaries.

\begin{figure}[tb]
\includegraphics[width=85mm,clip]{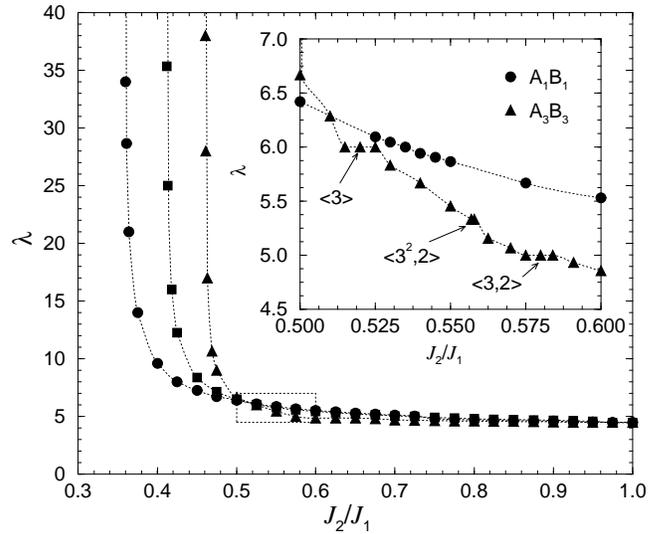}
\caption{The divergence of $\lambda$  the boundary with the ferromagnetic
phase. The black circles, squares, and triangles, respectively, correspond to
the selected points on the lines $A_1B_1$, $A_2B_2$, and $A_3B_3$. The inset
shows behavior of $\lambda$ around the commensurate phase $\langle3\rangle$.
The inset corresponds to the magnified area denoted by the dotted rectangle.}
\label{AB}
\end{figure}

\begin{figure}[tb]
\includegraphics[width=85mm,clip]{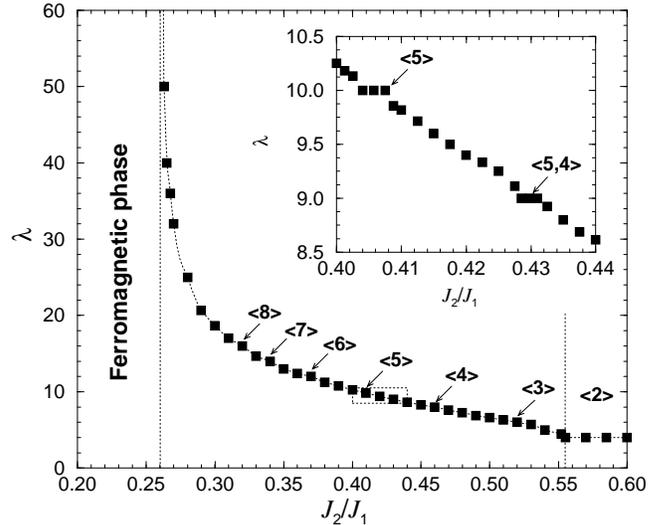}
\caption{The behavior of the wavelength $\lambda$ on the line $P_{\rm L}B_0$.
Several commensurate phases phases with the integer value of $\lambda$ are
labeled. The inset shows the area around the the commensurate phase
$\langle5\rangle$.}
\label{CD}
\end{figure}

In Fig.~\ref{AB} we plot the wavelength with respect to $J_2/J_1$ calculated
on the three lines $A_1B_1$, $A_2B_2$, and $A_3B_3$. The dotted line is
a guide for the eye to point out this structure. Near the boundary with the
ferromagnetic phase, the wavelength rapidly increases. This is contradictory
to the known results coming from the mean-field approximation. The mean-field
approximation yields the first-order transitions between ferromagnetic phase
and the individual commensurate phases on the boundary line (in details,
see Ref.~\cite{MFA}).

In the inset of Fig.~\ref{AB} we plot details of the
wavelength $\lambda$ in the vicinity of the commensurate phase
$\langle3\rangle$. We observed that the commensurate phases
$\langle3\rangle$, $\langle3^2,2\rangle$, and $\langle3,2\rangle$
'lock-in' at small regions of $J_2/J_1$ (on the line $A_3B_3$)
and the so-called "devil's stairs" behavior is observed~\cite{RepBak}.
On the contrary, the stairs-like structure is not visible at higher temperatures,
as seen on the line $A_1B_1$ near the paramagnetic boundary.

\begin{figure}[tb]
\includegraphics[width=85mm,clip]{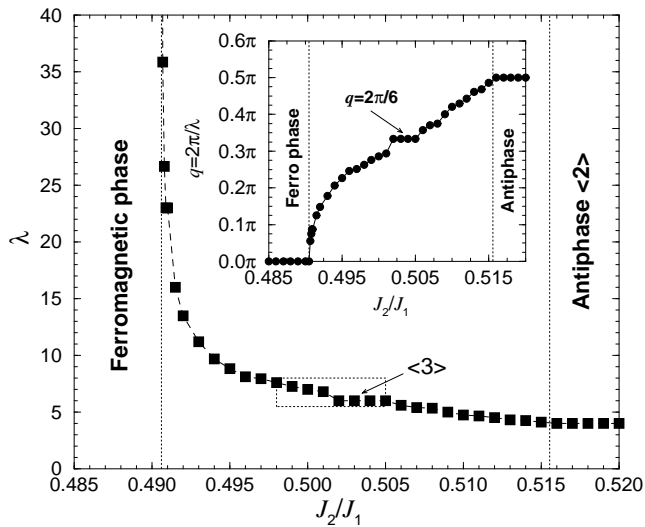}
\caption{The calculation of $\lambda$ on the line $A_4B_4$ at the temperature
$k_{\rm B}T/J_1=1.5$. The dotted rectangle borders an area shown in
Fig.~\ref{EFfine}. The inset illustrates the behavior of the corresponding
wave vector $q$.}
\label{EF}
\end{figure}

Figure~\ref{CD} shows $\lambda$ on the line from the point $B_0$ to the Lifshitz
point $P_{\rm L}$. The wavelength diverges toward the calculated Lifshitz
point at $J_2/J_1=0.26$. The stairs-like structure reveals if we zoom in 
on the phase diagram. For this reason, we selected an area depicted by
the dotted rectangle therein. The inset shows the magnified area with the
commensurate phases $\langle5\rangle$ and $\langle5,4\rangle$ where the
wavelength locks-in.

\subsection{Low temperature behavior}

To compare our results with analytical predictions, particularly, with the
low-temperature series expansions (LTSE)~\cite{FischerSelke}, we have selected
the line $A_4B_4$ which corresponds to the temperature $k_{\rm B}T/J_1=1.5$.
The computed data are plotted in Fig.~\ref{EF}. The commensurate phase
$\langle3\rangle$ ($\lambda=6$) locks-in and forms a well-visible plateau.
Note that the mean-field calculations~\cite{MFA} do not exhibit any phases
with $\lambda>6$ at $k_{\rm B}T/J_1\lesssim2$. It should be also noted that
both the mean-field approximation (at higher temperatures) and the LTSE
(at very low temperatures) do not result the spring of phases with
$\lambda>6$ from the multi-phase point $J_2/J_1=0.5$.

\begin{figure}[tp]
\includegraphics[width=85mm,clip]{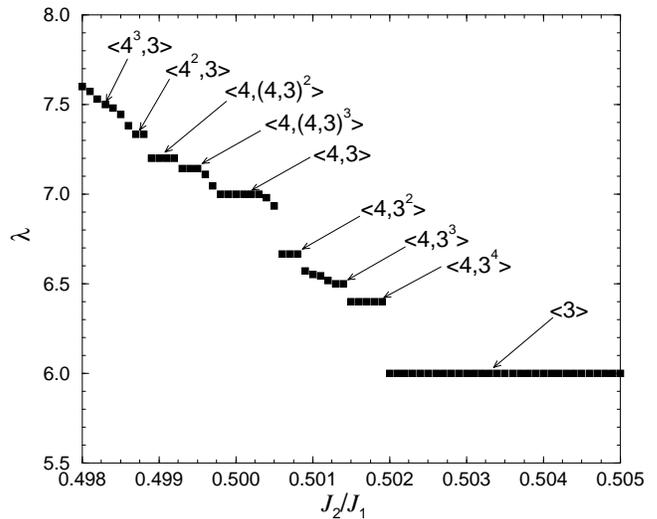}
\caption{The resulting devil's stairs with $\lambda\geq6$ observed on the line $A_4B_4$.}
\label{EFfine}
\end{figure}

Figure~\ref{EFfine} illustrates the stairs-like structure of $\lambda$
and corresponds to the magnified area shown by the dotted rectangle in
Fig.~\ref{EF}. The commensurate phases lock-in at rational values of
$\lambda$ and are separated by the high-order commensurate phases and
(possibly) the truly incommensurate ones. Several commensurate phases
are denoted above the stairs-like curve in Fig.~\ref{EFfine}.

The LTSE yields a spring of an infinity of the commensurate phases, such as
$\langle3\rangle$ and $\langle3,2^n\rangle$ for $n=1,2,3,\dots$, which separate
the ferromagnetic phase and the antiphase. The transition from the region of the
spin modulated phases ($\lambda>4$) to the antiphase ($\lambda=4$) does not
contradict to LTSE. The existence of
additional intermediate phases, $\langle3,2^n,3,2^{n+1} \rangle$ $n=1,2,3,\dots$,
at a higher temperatures was later reported by the LTSE.
Our results contain all these commensurate phases. Moreover, we found unpredicted
phases. For example, the transition between the phases $\langle3\rangle$ and
$\langle3,2\rangle$ is not of the first order as reported by LTSE. We found that
these two phases are separated by many commensurate phases, e.~g.,
$\langle3^n,2\rangle$ and $\langle3,(3,2)^{n+1}\rangle$ for
$n=1,2,3,\dots$ and the others of higher-orders as reported in Ref.~\cite{Surda}.
We obtained such rich spin modulated structure also for $\lambda>6$, see Fig.~\ref{EFfine}.

\begin{figure}[tp]
\includegraphics[width=85mm,clip]{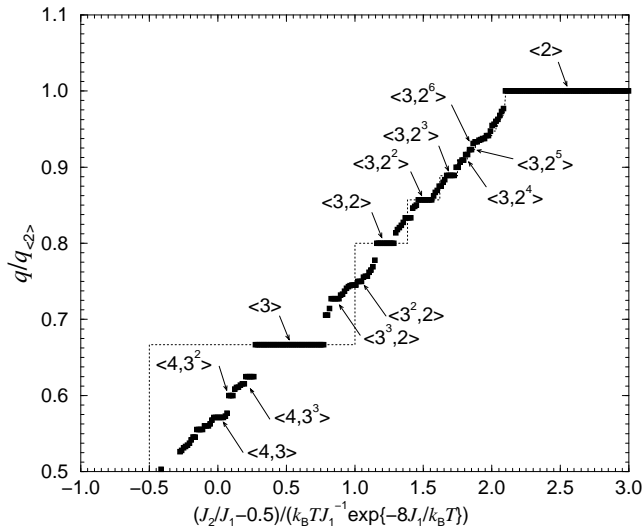}
\caption{Comparison of the numerical results at $k_{\rm B}T/J_1=1.5$
(the black squares) with the low-temperature series expansions represented
by the dashed stairs-like curve (in Ref.~\cite{FischerSelke}).}
\label{Cmp}
\end{figure}

In Fig.~\ref{Cmp} we depict our numerical results obtained at $k_{\rm B}
T/J_1=1.5$ and compare them with the results obtained by LTSE. The notation
$q_{\langle2\rangle}$ corresponds to the antiphase wave vector $q=\pi/2$.
Note that while the LTSE gives the first-order transition among individual
commensurate phases, our calculations yield subsequent stairs-like
structures among them. Moreover, the LTSE calculations do not yield
the commensurate phases with $\lambda>6$.

We, therefore, conjecture that the 'complete' devil's stairs structure
exists at intermediate temperatures. The complete devil's stairs
structure suggests there are no first-order transition~\cite{RepBak}.

Here, we summarize those commensurate phases which were obtained by the numerical
analysis of this model. Between two main commensurate phases $\langle p\rangle$
and $\langle p+1\rangle$, where $p=2,3,4,\dots$, new high-order commensurate
phases are present, such as $\langle p^{n-1},p+1\rangle$,
$\langle p,(p+1)^n\rangle$, $\langle p,(p,p+1)^{n-1}\rangle$,
and $\langle (p,p+1)^n,p+1\rangle$, with $n=2,3,4,\dots$
Subsequently, the following higher-order commensurate phases were found
$\langle (p^{n+1},p+1)^m,p^n,p+1\rangle$ and
$\langle (p,(p,p+1)^{n+1})^m,p,(p,p+1)^n\rangle$, for $m=2,3,4,\dots$
etc.

\section{Summary}

We applied the modified TPVA to the 3D ANNNI model and obtained the global
phase diagram. The location of the Lifshitz point agrees with calculations
performed by the high-temperature series expansions~\cite{Redner,Otmaa}
and the recent Monte Carlo calculations~\cite{Henkel}. The modulated phase
exhibits very complex structures. We found that (1) the commensurate phases
are substantially narrower than those reported so far, (2) the wavelength
of the spin modulated (commensurate) phases diverges at the boundary
with the ferromagnetic phase, (3) the commensurate phases merge at
low temperatures tending toward the multi-phase point $J_2/J_1=0.5$
and at low temperatures, the wavelengths with $\lambda>6$ are
obtained, and (4) many (possibly infinity) phases have been found
within the modulated phase, which have not yet been reported.

\section*{Acknowledgments}

A.~G. thanks A. \v{S}urda for an interesting discussion about the incommensurate phases
in the ANNNI model. This work has been partially supported by the Grant-in-Aid for
Scientific Research from Ministry of Education, Science, Sports and Culture (Grant
No.~09640462 and No.~11640376) and by the Slovak Grant Agency, VEGA No.~2/7201/21 and
2/3118/23. A.G. is also supported by Japan Society for the Promotion of Science (P01192).

\section*{Appendix}
\subsection*{Optimizing process for the local variational weights}

We briefly describe the optimizing process of finding out the local
variational weights in order to maximize Eq.~(\ref{vpf}). This optimization
is based on the self-consistent equation in the TPVA to achieve the minimum
of the free energy. Numerical details in the TPVA has been reported in Refs.~\cite{TPVA}.

In order to maximize the variational partition function in Eq.~(\ref{vpf}),
by a proper tuning of the local variational weights $V$, we define two
objects. One is the matrix object ${\cal B}$ that represents a punctured classical
system~\cite{Delgado} defined on the 2-layer spin system which corresponds
to the numerator $\langle\Psi|{\cal T}|\Psi\rangle$ of the variational partition
function. It is defined as
\begin{eqnarray}
\nonumber
{\cal B}_{i,0}\{\sigma|\bar\sigma\}&=&W^{\rm B}_{i,0}\{\sigma|\bar\sigma\}
\sum\limits_{\tilde{[\sigma]},\tilde{[\bar\sigma]}}\ \ \prod\limits_{k\neq i}
\ \ \prod\limits_{\ell\neq0}V_{k,\ell}\{\sigma\}\\
&\times&W^{\rm B}_{k,\ell}
\{\sigma|\bar\sigma\}V_{k,\ell}\{\bar\sigma\}.
\label{Bmat}
\end{eqnarray}
Analogously, the vector object ${\cal A}$ is the punctured system
defined on the 1-layer spin system,
\begin{equation}
{\cal A}_{i,0}\{\sigma\}=\sum\limits_{\tilde{[\sigma]}}\ \ \prod\limits_{k\neq i}
\ \ \prod\limits_{\ell\neq0}V_{k,\ell}\{\sigma\}V_{k,\ell}\{\sigma\}.
\label{Amat}
\end{equation}
The configuration sums in Eqs.~(\ref{Bmat}) and (\ref{Amat}) are taken over all
the spin variables $\sigma$ except for the 6 ones at the center of the system.
In particular, except for
\begin{equation}
\{\sigma\}=(\sigma_{i,0}\ \sigma_{i+1,0}\ \sigma_{i+2,0}
\ \sigma_{i,1}\ \sigma_{i+1,1}\ \sigma_{i+2,1})
\end{equation}
and analogously for $\{\bar\sigma$\} in Eq.~(\ref{Bmat}). The notations
$\prod_{k\neq i}\prod_{\ell\neq0}$
exclude $V_{i,0}\{\sigma\}$ and $V_{i,0}\{\bar\sigma\}$ from the product.
Having defined these two objects, the variational partition function can
be transformed into the expression,
\begin{equation}
\lambda_{\rm var}=\frac{\sum\limits_{\{\sigma\},\{\bar\sigma\}}V_{i,0}\{\sigma\}
{\cal B}_{i,0}\{\sigma|\bar\sigma\}V_{i,0}\{\bar\sigma\}}
{\sum\limits_{\{\sigma\}}V_{i,0}\{\sigma\}{\cal A}_{i,0}\{\sigma\}
V_{i,0}\{\sigma\}}.
\label{pfab}
\end{equation}

Now, consider a variation of $\lambda_{\rm var}$ with respect to variations
of the local variational weights
\begin{equation}
\frac{\delta\lambda_{\rm var}}{\delta\Psi}\equiv\sum\limits_{i,j}
\frac{\delta\lambda_{\rm var}}{\delta V_{i,j}}.
\end{equation}
Carrying out the extremal condition, $\delta\lambda/\delta V_{i,j}=0$,
the self-consistent equation for the local variational weights $V_{i,j}$
is then obtained
\begin{equation}
V_{i,0}^{\rm new}\{\sigma\}=
\sum\limits_{\{\bar\sigma\}}\frac{{\cal B}_{i,0}\{\sigma|\bar\sigma\}}
{{\cal A}_{i,0}\{\sigma\}}V_{i,0}\{\bar\sigma\}.
\label{sce}
\end{equation}
The improvement of $V$ is performed as
\begin{equation}
V_{i,0}\{\sigma\}=V_{i,0}\{\sigma\}+\varepsilon V_{i,0}^{\rm new}\{\sigma\}
\label{impr}
\end{equation}
through 64-parameter local parameter adjust.
The self-consistent relation, Eq.~(\ref{sce}),
is a non-linear equation since ${\cal B}_{i,0}$ and ${\cal A}_{i,0}$
themselves depend on $V$. The convergence parameter $\varepsilon$ controls
the rate at which the improvement process of $V$ is performed.

Consequently, we compute the free energy per strip
\begin{equation}
{\cal F}_{\rm new}=-k_{\rm B}T\ln\lambda_{\rm var}
\label{feng}
\end{equation}
and compare with the free energy ${\cal F}_{\rm old}$ calculated with the
previous $V$.

\subsection*{Efficiency of the algorithm}

After the trial state $|\Psi\rangle$ is optimized, we calculate the spontaneous
magnetization at a site
\begin{equation}
\langle\sigma_{i,j}\rangle=\frac{\langle\Psi|\sigma_{i,j}|\Psi\rangle}
{\langle\Psi|\Psi\rangle}.
\label{mag}
\end{equation}
Since the competing interactions exist only along the $x$ direction, the system
is translation invariant with respect to the $y$ and $z$ directions.
Therefore, the spontaneous magnetization $\langle\sigma_{i,j}\rangle$ is
independent on $j$ and we used $\langle\sigma_i\rangle$ instead.

\begin{table}[t]
\caption {The critical temperature $T_{\rm c}$ for the 3D Ising model ($J_2\!=\!0$).
We calculate the relative errors $\epsilon$ with respect to $T_{\rm c}$ obtained
by Monte Carlo simulations~\cite{MC}.}
\label{t}
\begin{ruledtabular}
\begin{tabular*}{\hsize}{
l@{\extracolsep{0ptplus1fil}}
c@{\extracolsep{0ptplus1fil}}r}
{\bf Numerical method} & $T_{\rm c}$ & $\epsilon$/[\%] \\
\colrule
Mean-field approximation~\cite{SelkeRev} & 6.000 & 33.0 \\
Kramers-Wannier approximation~\cite{KWA} & 4.587 & 1.7 \\
TPVA with 16 parameters~\cite{TPVA} & 4.570 & 1.3 \\
TPVA with 64 parameters & 4.554 & 0.9 \\
Monte Carlo simulations & 4.512 & --- \\
\end{tabular*}
\end{ruledtabular}
\end{table}

In order to estimate the numerical accuracy of the improved TPVA, we compare the
calculation of the critical temperature $T_{\rm c}$ in the pure Ising model, i.~e.,
when $J_2=0$, with other numerical methods. Table~II summarizes the obtained
$T_{\rm c}$. It is obvious that the mean-field approximation overestimates
$T_{\rm c}$ and does not give reliable results near the phase boundaries.
The improved TPVA with the 64 variational parameters results better
$T_{\rm c}$ than the original TPVA with 16 parameters~\cite{TPVA,rem2}.

We set up the convergence parameter $|\varepsilon|=10^{-2}$.
Assuming any $|\varepsilon|\lesssim10^{-2}$ is sufficient for the most cases.

\begin{figure}[tb]
\includegraphics[width=85mm,clip]{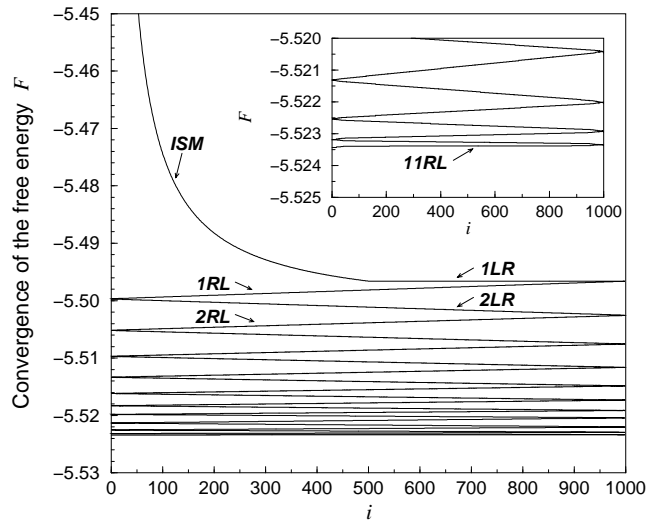}
\caption{Typical convergence process of the free energy ${\cal F}$ for the lattice
size of $1001\times\infty\times\infty$. The inset shows a detailed view
near the vicinity of the free energy minimum up to the 11$^{\rm th}$ sweep (11RL).}
\label{fe}
\end{figure}

In Fig.~\ref{fe} we illustrate an example which demonstrates the systematic decay
of the free energy during the DMRG sweeping process until it finally converges.
After the DMRG infinite system method (ISM) is finished, the first left-right sweep
(1LR) proceeds followed by the first right-left sweep (1RL) and so on. Each step of
the finite system method decreases the free energy until its minimum is reached.

\end{document}